\documentstyle[aps,preprint]{revtex}

\begin{document}

\title{
Effects of phonon-phonon coupling on low-lying states in neutron-rich Sn isotopes.}

\author{A.P. Severyukhin}
\address{Bogoliubov Laboratory of Theoretical Physics, Joint Institute for
Nuclear Research, 141980 Dubna, Moscow region, Russia}
\author{V.V. Voronov}
\address{Bogoliubov Laboratory of Theoretical Physics, Joint Institute for
Nuclear Research, 141980 Dubna, Moscow region, Russia}
\author{Nguyen Van Giai}
\address{Institut de Physique
Nucl\'eaire, CNRS-IN2P3, Universit\'e Paris-Sud, F-91406 Orsay Cedex, France}

\maketitle
\begin{abstract}

Starting from an effective Skyrme interaction we present a method
to take into account the coupling between one-
and two-phonon terms in the wave functions of excited states.
The approach is a development of a finite rank separable approximation
for the quasiparticle RPA calculations proposed in our previous work.
The influence of the phonon-phonon coupling on energies and transition probabilities
for the low-lying quadrupole and octupole states in the neutron-rich Sn isotopes is studied.
\end{abstract}
\pacs{PACS numbers: 21.60.Jz, 24.30.Cz, 21.30.Fe, 21.65.+f}

\section{Introduction}

The experimental and theoretical studies of properties of the excited states in nuclei far from the $\beta$-stability line
%are very popular now.
are presently the object of very intensive activity.
The random phase approximation (RPA) \cite{R70,BM75,Schuck,solo} is a well-known and
successful way to treat nuclear vibrational excitations.
Using the Gogny's \cite{gogny} or Skyrme-type \cite{vau72} effective nucleon-nucleon interactions
the most consistent models can describe the ground states in the
framework of the Hartree-Fock (HF) and Hartree-Fock-Bogoliubov (HFB)
approximations
and the excited states within the RPA and quasiparticle RPA (QRPA).
Such models are quite successful not only to reproduce the nuclear ground state
properties \cite{floc78,doba96}, but also to describe the main features of
nuclear excitations in closed-shell \cite{colo,colo1} and
open-shell nuclei \cite{KG00,colo2,colo3,KSGG03}. In the latter case the pairing correlations are very important.

Due to the anharmonicity of vibrations there is a
coupling between one-phonon and more complex states \cite{BM75,solo}
and the complexity of calculations beyond standard RPA or QRPA increases rapidly
with the size of the configuration space, so one has to work within limited spaces.
Making use of separable forces one can perform calculations of nuclear characteristics
in very large configuration spaces since there is no need to diagonalize matrices
whose dimensions grow with the size of configuration space.
For example, the well-known quasiparticle-phonon model (QPM) \cite{solo}
can do very detailed predictions for nuclei away from closed shells \cite{gsv}, but it is very difficult to extrapolate
the phenomenalogical parameters of the nuclear hamiltonian to new regions of nuclei.

That is why a finite rank approximation
for the particle--hole (p-h) interaction resulting from
the Skyrme forces has been suggested in our previous work\cite{gsv98}.
Thus, the self-consistent mean field can be calculated
with the original Skyrme interaction whereas the RPA
solutions would be obtained with the finite rank approximation to the
p-h~matrix elements. It was found that the finite rank approximation can reproduce reasonably well
the dipole and quadrupole strength distributions in Ar isotopes.
Alternative schemes to factorize the p-h interaction
were considered in \cite{suz81,sar99,nest02}.

Recently, the finite rank approximation for p-h interactions
of Skyrme type has been generalized to take into account the pairing correlations \cite{ssvg02}.
%The quasiparticle RPA (QRPA)
The QRPA was used to describe characteristics of the low-lying
$2^+$ and $3^-$ states and giant resonances in nuclei with very different
mass numbers \cite{ssvg02,svsg03}. It was found that there is room for the phonon-phonon coupling effects
in many cases. The first calculation to estimate this effect has been done for $^{112}$Sn in \cite{svsg103}.

In the present work, we extend our approach to take into account
the coupling between the one- and two-phonon terms in the wave functions
of excited states.
As an application of the method we present results
%of calculations
for low-lying $2^+$ and $3^-$ states
in neutron-rich Sn isotopes and compare them with recent experimental data \cite{hribf} and
other calculations \cite{tens02,colo03,giam03}.

This paper is organized as follows: in Sec. II we sketch our method
allowing to consider effects of the phonon-phonon coupling.
In Sec. III we discuss details of calculations and show how
this approach can be applied to treat the low-lying collective states.
Results of calculations for properties of the quadrupole and
octupole states in $^{124-134}$Sn isotopes are given in Sec.IV.
Conclusions are drawn in Section V.

\section{Method of calculations}
\subsection{The model hamiltonian and QRPA}
We start from the effective Skyrme interaction\cite{vau72}
and  use the notation of Ref.\cite{sg81} containing explicit
density dependence and all spin-exchange terms.
The single-particle spectrum is calculated within the HF method.
The continuous part of the single-particle spectrum is
discretized by diagonalizing the HF hamiltonian on
a harmonic oscillator basis\cite{BG77}.
The p-h residual interaction $\tilde V_{res}$
corresponding to the Skyrme force and including both direct and exchange
terms can be obtained as the second derivative of the energy density
functional with respect to the density\cite{ber75}.
Following our previous papers\cite{gsv98} we  simplify $\tilde V_{res}$ by
approximating it by its Landau-Migdal form.
For Skyrme interactions all Landau parameters $F_l, G_l, F^{'}_l, G^{'}_l$
with $l > 1$ are zero. Here, we keep only the $l=0$ terms
in $V_{res}$ and in the coordinate
representation one can write it in the following form:
\begin{eqnarray}
V_{res}({\bf r}_1,{\bf r}_2)=N_0^{-1}\left[ F_0(r_1)+G_0(r_1)
{\bf \sigma}_1\cdot{\bf \sigma}_2+(F_0^{^{\prime
}}(r_1)+G_0^{^{\prime }}(r_1){\bf \sigma }_1\cdot{\bf \sigma}_2){\bf
\tau }_1\cdot{\bf \tau }_2\right] \delta ({\bf r}_1-{\bf r
}_2)  \label{eq2}
\end{eqnarray}

where ${\bf \sigma}_i$ and ${\bf \tau}_i$ are the
spin and isospin operators,
and $N_0 = 2k_Fm^{*}/\pi^2\hbar^2$ with $k_F$ and $m^{*}$ standing for the
Fermi momentum and nucleon effective mass.
The expressions for
$F_0, G_0, F^{'}_0, G^{'}_0$ in terms of the Skyrme force parameters can
be found in Ref.\cite{sg81}. Because of
the density dependence of the interaction the Landau parameters of
Eq.(\ref{eq2}) are functions of the coordinate ${\bf r}$.

In what follows we use the second quantized representation
and $V_{res}$ can be written as:

\begin{eqnarray}
\hat V_{res} & = & \frac 12\sum_{1234}V_{1234}:a_1^{+}a_2^{+}a_4 a_3:
%\label{eq3}
\end{eqnarray}

where $a^+_1$ ($a_1$) is the particle creation (annihilation) operator
and $1$ denotes the quantum numbers $(n_1l_1j_1m_1)$,

\begin{eqnarray}
V_{1234} = \int \phi^*_1({\bf r}_1)\phi^*_2({\bf r}_2)
V_{res}({\bf r}_1,{\bf r}_2)\phi_3({\bf r}_1)
\phi_4({\bf r}_2) {\bf dr}_1{\bf dr}_2 .
%\label{eq4}
\end{eqnarray}

After integrating over the angular variables one needs
to calculate the radial integrals.
%\cite{gsv98,ssvg02}.
As it is shown in \cite{gsv98,ssvg02} the radial integrals can be calculated
accurately by choosing a large enough cutoff radius $R$
and using a $N$-point integration Gauss formula with abscissas ${r_k}$ and weights ${w_k}$.
Thus, the two-body matrix element
%residual interaction
is a sum of $N$ separable terms, i.e., the residual interaction takes the
form of a rank $N$ separable interaction.

%So we
We employ a hamiltonian
including an average HF field,
pairing interactions, the isoscalar and
isovector particle--hole (p--h)
residual forces in a finite rank separable form \cite{ssvg02}:

\begin{eqnarray}
H &=&\sum\limits_\tau \left( \left. \sum\limits_{jm}\right. ^\tau
(E_j-\lambda _\tau )a_{jm}^{\dagger }a_{jm}-\frac 14V_\tau
^{(0)}:P_0^{\dagger }\,(\tau )P_0\,(\tau ):\right)+ \hat V_{res}
\label{eq9},
\end{eqnarray}

where

\begin{equation}
P_0^{+}\,(\tau )=\,\left. \sum_{jm}\right. ^\tau
(-1)^{j-m}a_{jm}^{+}a_{j-m}^{+}.
\end{equation}

We sum over the proton($p$) and neutron($n$) indexes and the notation $
\{\tau =(n,p)\}$ is used. A change $\tau \leftrightarrow -\tau $ means a
change $p\leftrightarrow n$.
The single-particle states are specified by the
quantum numbers $(jm)$, $E_j$ are the single-particle energies,
$\lambda_\tau $ the chemical potentials. $V_\tau ^{(0)}$  is the
interaction strength in the particle-particle channel.
The hamiltonian (\ref{eq9}) has the same form as the QPM
hamiltonian with $N $ separable terms \cite{solo,sol89}, but the
single-particle spectrum and parameters of the p-h residual interaction are
calculated making use of the Skyrme forces.

In what follows we work in the quasiparticle  representation defined by
the canonical Bogoliubov transformation:

\begin{equation}
a_{jm}^{+}\,=\,u_j\alpha _{jm}^{+}\,+\,(-1)^{j-m}v_j\alpha _{j-m}.
\label{B}
\end{equation}

The hamiltonian (\ref{eq9}) can be represented in terms of bifermion
quasiparticle operators and their conjugates \cite{solo}:

\begin{equation}
B(jj^{^{\prime }};\lambda \mu )\,=\,\sum_{mm^{^{\prime }}}(-1)^{j^{^{\prime
}}+m{^{\prime }}}\langle jmj^{^{\prime }}m^{^{\prime }}\mid \lambda \mu
\rangle \alpha _{jm}^{+}\alpha _{j^{^{\prime }}-m^{^{\prime }}},
\end{equation}
\begin{equation}
A^{+}(jj^{^{\prime }};\lambda \mu )\,=\,\sum_{mm^{^{\prime }}}\langle
jmj^{^{\prime }}m^{^{\prime }}\mid \lambda \mu \rangle \alpha
_{jm}^{+}\alpha _{j^{^{\prime }}m^{^{\prime }}}^{+}.
\end{equation}

We introduce the phonon creation operators

\begin{equation}
Q_{\lambda \mu i}^{+}\,=\,\frac 12\sum_{jj^{^{\prime }}}\left( X
_{jj^{^{\prime }}}^{\lambda i}\,A^{+}(jj^{^{\prime }};\lambda \mu
)-(-1)^{\lambda -\mu }Y _{jj^{^{\prime }}}^{\lambda i}\,A(jj^{^{\prime
}};\lambda -\mu )\right).
\end{equation}

where the index $\lambda $ denotes total angular momentum and $\mu $ is
its z-projection in the laboratory system.
One assumes that the ground state  is the QRPA phonon vacuum
$\mid 0\rangle $,\\ i.e. $Q_{\lambda \mu i}\mid 0\rangle\,=0$.
We define the excited states for this approximation by
$Q_{\lambda\mu i}^{+}\mid0\rangle$.
The quasiparticle energies ($\varepsilon _j$),
the chemical potentials ($\lambda_\tau $), the energy gap and the coefficients
$u$,$v$ of the  Bogoliubov transformations
(\ref{B}) are determined
from  the BCS equations with the single-particle spectrum that is calculated
within the HF method with the effective Skyrme interaction. Making use
of the linearized equation-of-motion approach \cite{R70}:

\begin{equation}
\langle 0|\left[ \delta Q_{\lambda \mu i},\left[ H,Q_{\lambda \mu
i}^{+}\right] \right] \mid 0\rangle =\omega _{_{\lambda i}}\langle 0|\left[
\delta Q_{\lambda \mu i},Q_{\lambda \mu i}^{+}\right] \mid 0\rangle,
%\label{eq11}
\end{equation}
with the normalization condition:
\begin{equation}
\langle 0\mid \,[Q_{\lambda \mu i},Q_{\lambda \mu i^{^{\prime }}}^{+}]\,\mid
0\rangle \,=\delta _{ii^{^{\prime }}},
\label{eq12}
\end{equation}

one can get the QRPA equations \cite{Schuck,solo}:

\begin{equation}
\label{eq14}
\left(
\begin{tabular}{ll}
${\cal A}$ & ${\cal B}$ \\
${- \cal B}$ & ${- \cal A}$%
\end{tabular}
\right) \left(
\begin{tabular}{l}
$ X $ \\
$ Y $%
\end{tabular}
\right) =w \left(
\begin{tabular}{l}
$ X $ \\
$ Y $%
\end{tabular}
\right).
\end{equation}

In QRPA problems there appear two types of interaction matrix elements,
the $A^{(\lambda)}_{(j_1j_1^{\prime})(j_2j_2^{\prime})}$
matrix related to forward-going graphs and the
$B^{(\lambda)}_{(j_1j_1^{\prime})_{\tau}(j_2j_2^{\prime})_{q\tau}}$
matrix related to backward-going graphs.
Solutions of this set of linear equations yield the eigen-energies
and the amplitudes $X,Y$ of the excited states.
The dimension of the matrices ${\cal A}, {\cal B}$ is the space size of
the two-quasiparticle configurations. For our case expressions
for ${\cal A}, {\cal B}$ and $X,Y$ are given in \cite{ssvg02}.
Using the finite rank approximation we need to invert a matrix of
dimension $4N \times 4N$ independently of
the configuration space size \cite{gsv98,ssvg02}.
Therefore, this approach enables one to reduce remarkably the dimensions of
the matrices that must be inverted to perform structure calculations
in very large configuration spaces.

\subsection{Phonon-phonon coupling}

Our calculations \cite{ssvg02} show that, for the normal parity states one can neglect
the spin-multipole terms of the p-h residual interaction (\ref{eq2}).
Using the completeness and orthogonality conditions for the
phonon operators one can express bifermion operators
$A^{+}(jj^{^{\prime }};\lambda \mu )$ and  $A(jj^{^{\prime }};\lambda
\mu)$  through the phonon ones and
the initial hamiltonian (\ref{eq9}) can be rewritten in terms of
quasiparticle and phonon operators in the following form:

\begin{equation}
H=\,h_0\,+\,h_{QQ}\,+\,h_{QB}
\label{coup1}
\end{equation}
\begin{equation}
h_0=\,\sum_{jm}\,\varepsilon _j\,\alpha _{jm}^{+}\,\alpha_{jm}
%\label{}
\end{equation}
\begin{equation}
h_{QQ}=-\frac 14\sum_{\lambda \mu ii^{^{\prime }}\tau }
W^{\lambda ii^{\prime }}\left( \tau \right)
Q_{\lambda \mu i}^{+}Q_{\lambda \mu i^{^{\prime}}}
%\label{}
\end{equation}
\begin{eqnarray}
h_{QB} &=&-\frac 1{2}\sum_{\lambda \mu i\tau }\left.
\sum_{jj^{^{\prime }}}\right. ^\tau \Gamma_{jj'}^{\lambda i}\left( \tau \right)\left( (-)^{\lambda -\mu }Q_{\lambda \mu
i}^{+} + Q_{\lambda -\mu i}\right) B(jj^{^{\prime }};\lambda -\mu )+h.c.
%\label{}
\end{eqnarray}

The coefficients $W$, $\Gamma$ of the hamiltonian (\ref{coup1}) are sums
of N combinations of phonon amplitudes,
the Landau parameters, the reduced matrix element of the spherical harmonics and
radial parts of the HF single-particle wave function (see Appendix A).
It is worth to point out that the term $h_{QB}$ is responsible for the mixing of the configurations and,
therefore, for the description of many characteristics of the excited states
of even--even nuclei \cite{solo}.

To take into account the mixing of the configurations
in the simplest case one can write the wave functions of excited
states as:

\begin{equation}
\Psi _\nu (\lambda \mu )=\{\sum_iR_i(\lambda \nu )Q_{\lambda \mu
i}^{+}+\sum_{\lambda _1i_1\lambda _2i_2}P_{\lambda _2i_2}^{\lambda
_1i_1}(\lambda \nu )\left[ Q_{\lambda _1\mu _1i_1}^{+}Q_{\lambda _2\mu
_2i_2}^{+}\right] _{\lambda \mu }\}|0\rangle
\label{wf2ph}
\end{equation}

with the normalization condition:

\begin{equation}
\sum\limits_iR_i^2(J\nu)+ 2\sum_{\lambda _1i_1 \lambda _2i_2}
(P_{\lambda _2i_2}^{\lambda _1i_1}(J\nu))^2=1
\label{norma2ph}
\end{equation}

Using the variational principle in the form:

\begin{equation}
\delta {\ }\left( {\langle \,\Psi _\nu (\lambda \mu )\,|H\mid \,\Psi _\nu
(\lambda \mu )\rangle \,-\,E}_\nu \left( \langle {\,\Psi _\nu (\lambda \mu
)\,|\,\Psi _\nu (\lambda \mu )\rangle \,-\,1}\right) \right) \,=\,0,
%\label{}
\end{equation}

one obtains a set of linear equations for
the unknown amplitudes $R_i(J\nu)$ and $P_{\lambda_2i_2}^{\lambda_1i_1}(J\nu )$:

\begin{equation}
(\omega _{Ji}-E_\nu )R_i(J\nu )+\sum_{\lambda _1i_1 \lambda_2i_2} U_{\lambda _2i_2}^{\lambda _1i_1}(J\nu)
P_{\lambda_2i_2}^{\lambda _1i_1}(J\nu )=0
\label{2pheq1}
\end{equation}
\begin{equation}
\sum\limits_iU_{\lambda _2i_2}^{\lambda _1i_1}(Ji)R_i(J\nu ) +
2(\omega _{\lambda _1i_1}+\omega _{\lambda _2i_2}-E_\nu )
P_{\lambda _2i_2}^{\lambda _1i_1}(J\nu)=0
\label{2pheq2}
\end{equation}

$U_{\lambda _2i_2}^{\lambda _1i_1}(J i)$ is the matrix element
coupling one- and two-phonon configurations \cite{solo,vs83}:

\begin{equation}
U_{\lambda _2i_2}^{\lambda _1i_1}(J i)=
\langle 0| Q_{J i } h_{QB} \left[ Q_{\lambda _1i_1}^{+}Q_{\lambda
_2i_2}^{+}\right] _J |0 \rangle.
%\label{}
\end{equation}

The expression of $U_{\lambda _2i_2}^{\lambda _1i_1}(J i)$
is given in Appendix B.
The number of linear equations (\ref{2pheq1}), (\ref{2pheq2}) equals the number of one-
and two-phonon configurations included in the wave function (\ref{wf2ph}).

The energies of  excited states $E_\nu$ are solutions of the secular equation

\begin{equation}
F(E_\nu )\equiv det\left| (\omega _{\lambda i}-E_\nu )\delta _{ii^{\prime
}}-\frac 12\sum_{\lambda _1i_1,\lambda _2i_2}\frac{U_{\lambda
_2i_2}^{\lambda _1i_1}(\lambda i)U_{\lambda _2i_2}^{\lambda _1i_1}(\lambda
i^{\prime })}{\omega
_{\lambda _1i_1}+\omega _{\lambda _2i_2}-E_\nu }\right| =0,
%\label{}
\end{equation}

where the rank of the determinant equals the number of the
one-phonon configurations.
Using Eqs.(\ref{2pheq1}), (\ref{2pheq2}) and the normalization condition (\ref{norma2ph}),
one can find the amplitudes
$R_i(J\nu)$ and  $P_{\lambda _2i_2}^{\lambda_1i_1}(J\nu )$.

It is necessary to point out that the equations derived above have the same form as
the basic QPM equations \cite{solo,vs83},
but the single-particle spectrum and the p-h residual interaction are determined making use of
the Skyrme interactions.

\section{Details of calculations}

We apply the present approach to study characteristics of the low-lying
vibrational states in the neutron-rich Sn isotopes. In this paper
we use the parametrization SLy4 \cite{sly4} of the Skyrme interaction.
This parametrization was proposed to describe isotopic properties of nuclei from
the $\beta$-stability line to the drip lines.
Spherical symmetry is assumed for the HF ground states.

The pairing constants $V^0_{\tau}$ are fixed to reproduce the odd-even mass
difference of neighboring nuclei.
It is well known \cite{KG00,colo2} that the constant gap approximation leads to
an overestimating of occupation probabilities for subshells that are far
from the Fermi level and it is necessary to introduce a cut-off in the
single-particle space. Above this cut-off subshells don't participate in
the pairing effect. In our calculations we choose the BCS subspace
to include all subshells lying below 5 MeV.

In order to perform QRPA calculations, the single-particle continuum is
discretized \cite{BG77} by diagonalizing the HF hamiltonian on a basis
of twelve harmonic oscillator shells and cutting off the single-particle
spectra at the energy of 100 MeV.
This is sufficient to exhaust practically all the
energy-weighted sum rule.

The Landau parameters $F_0$, $G_0$, $F_0^{'}$, $G_0^{'}$ expressed in terms of
the Skyrme force parameters \cite{sg81} depend on $k_F$.
As it is pointed out in our previous works \cite{gsv98,ssvg02} one needs to
adopt some effective value for $k_F$ to give an accurate representation
of the original p-h Skyrme interaction.
For the present calculations we use the nuclear matter value for $k_F$.

Our previous investigations \cite{ssvg02} enable us to conclude that $N$=45
for the rank of our separable approximation is enough for
multipolarities $\lambda \le 3 $ in nuclei with $A\le 208$.
Increasing $N$, for example, up to $N$=60 in $^{208}$Pb changes results
for energies and transition probabilities not more than by 1\%.
Our calculations show that, for the
%normal
natural parity states one can neglect
the spin-multipole interactions and this reduces by a factor 2
the total matrix dimension, i.e., the matrix dimensions never exceed
$2N \times 2N$ independently of the configuration space size \cite{gsv98,ssvg02}.

The two-phonon configurations of the wave function (\ref{wf2ph})
are constructed from
%normal
natural parity phonons with multipolarities $\lambda = 2,3,4,5$.
All one-phonon configurations with energies below 8 MeV for $^{124-130,134}$Sn
and 10 MeV for $^{132}$Sn are included in the the wave function (\ref{wf2ph}).
The cut-off in the space of the two-phonon configurations is 21 MeV.
An extension of the space for one- and two-phonon configurations does not change results
for energies and transition probabilities practically.

\section{Results of calculations}

As an application of the method we investigate effects of the phonon-phonon coupling on
energies and transition probabilities to $2^+_1$ and  $3^-_1$ states
in $^{124-134}$Sn.

Results of our calculations for the $2^+_1$ energies and transition probabilities $B(E2)$
are compared with experimental data \cite{hribf,Ram01}
in Table I.
Columns "QRPA" and "2PH" give values calculated within the QRPA and
taking into account the phonon-phonon coupling, respectively.

As it is seen from  Table I there is a remarkable increase of
the $2^+_1$ energy and $B(E2\uparrow)$ in
$^{132}$Sn in comparison with those in $^{130,134}$Sn.
Such a behaviour of $B(E2\uparrow)$ is related
with the proportion between the QRPA amplitudes for neutrons and protons in Sn isotopes.
The neutron amplitudes are dominant in all Sn isotopes and the contribution of the main neutron
configuration $\{1h_{11/2},1h_{11/2}\}$ increases from  81.2\% in $^{124}$Sn to 92.8\% in $^{130}$Sn
when neutrons fill the subshell $1h_{11/2}$. At the same time the contribution of
the main proton configuration $\{2d_{5/2},1g_{9/2}\}$ is decreasing from 9.3\% in $^{124}$Sn to 3.9\% in $^{130}$Sn.
The closure of the neutron subshell $1h_{11/2}$ in $^{132}$Sn leads to the vanishing
of the neutron paring.
The energy of the first neutron two-quasiparticle pole $\{2f_{7/2},1h_{11/2}\}$ in $^{132}$Sn
is greater than energies of the first poles in $^{130,134}$Sn and the contribution of the
$\{2f_{7/2},1h_{11/2}\}$ configuration in the doubly magic $^{132}$Sn is about 61\%.
Furthermore, the first pole in $^{132}$Sn is closer to the proton poles.
This means that the contribution of the proton two-quasiparticle configurations
is greater than those in the neighbouring isotopes and as a result the main proton configuration $\{2d_{5/2},1g_{9/2}\}$
in $^{132}$Sn exhausts about 33\%. In $^{134}$Sn the leading contribution (about 99\%) comes from the neutron
configuration $\{2f_{7/2},2f_{7/2}\}$ and as a result the $B(E2)$ value is reduced.
Such a behaviour  of the  $2^+_1$ energies and $B(E2)$ values
in the neutron-rich Sn
isotopes reflects the shell structure in this region .
 It is worth to mention that the first prediction of the anomalous behaviour of $2^+$ excitations around
$^{132}$Sn based on the QRPA calculations with a separable quadrupole-plus-pairing hamiltonian has been done
in \cite{tens02}. Other QRPA calculations with Skyrme \cite{colo3} and Gogny \cite{giam03} forces give similar
results for Sn isotopes.

One can see from Table I that the inclusion of the two-phonon terms results in a decrease of the energies
and a reduction of transition probabilities.
Note that  the effect of the two-phonon configurations is important for the energies and
this effect becomes weak in $^{132}$Sn.  There is some overestimate of the energies
for the QRPA calculations  and taking into account of the
two-phonon terms improves the description of
the $2^+_1$ energies . The reduction of the $B(E2)$ values is small in most cases due to the crucial contribution of
the one-phonon configuration in the wave function structure.

Results of our calculations for the $3^-_1$ energies and the
transition probabilities $B(E3)$ compared to experimental data  \cite{Sp02}
are shown in Table II. As for the quadrupole excitations the influence of coupling between one-
and two-phonon terms in the wave functions of the $3^-_1$ states leads to
the decrease of the energies and the reduction of transition probabilities.
In spite of the fact that the $3^-_1$ states have strong collectivity and many two-quasiparticle configurations
give a contribution in the QRPA wave functions in Sn isotopes the phonon-phonon coupling is not very strong in this case.
Our calculation shows that the main reason is the smallness of
the matrix elements coupling the one-phonon configuration $\{3^-_1\}$ and the
two-phonon configuration $\{2^+_1; 3^-_1\}$ ($U_{3^-_1}^{2^+_1}(3^-_1)$).
As a result the decrease of the $3^-_1$ energies is about 10\%.  In the present paper we neglect
the p-p channel that can be important for collective phonons and
can reduce the collectivity of
states \cite{solo,kubo96}. This can give an additional lowering of energies  and transition probabilities , but
this is not the case for $^{132}$Sn. It is worth to mention that experimental data for $3^-_1$ states in the
neutron-rich Sn isotopes are very scarce.

An additional information  about the structure of the first $2^+, 3^-$
states can be extracted by looking at the ratio of the multipole
transition matrix elements $M_n/M_p$ that depends on the relative
contributions of the proton and neutron configurations. In the framework
of the collective model for isoscalar excitations this ratio is
equal to $M_n/M_p=N/Z$ and any deviation from this value can indicate
an isovector character of the state. The $M_n/M_p$ ratio can be determined
experimentally by using different external probes \cite{Ber83,Ken92,Jew99}.
Our calculated values for the $M_n/M_p$ ratios for the $2^+_1$ and $3^-_1$ states
are shown in Table III. The calculated $M_n/M_p$ ratios are rather close to $N/Z$ except
$2^+_1$ in $^{134}$Sn.
It is worth to note that the deviation of the ratio
for $2^+_1$ state in $^{132}$Sn correlates with the increase of the contribution
of the proton two-quasiparticle configurations.

\section{Conclusions}

A finite rank separable approximation for the QRPA
calculations with Skyrme interactions that was proposed in
our previous work is extended to take into account the coupling between one-
and two-phonon terms in the wave functions of excited states.
The suggested approach enables one to reduce remarkably the dimensions of the matrices
that must be diagonalized to perform structure calculations in very large
configuration spaces.
As an application of the method we have studied the behavior of the energies and
transition probabilities to $2^{+}_{1}$ and $3^{-}_{1}$ states in $^{124-134}$Sn.
The inclusion of the two-phonon configurations results in a decrease of the energies and
a reduction of transition probabilities.
It is shown that there is some overestimate of $2^{+}_{1}$ energies
for the QRPA calculations and the effect of the two-phonon configurations is important,
but this effect decreases in $^{132}$Sn. The inclusion of the two-phonon terms does not change
the effect of a remarkable increase of the
%calculated within
QRPA value of $B(E2; 0^{+} \rightarrow 2^{+}_{1})$
for the doubly-closed shell nucleus $^{132}$Sn in comparison with its neighbors.
A systematical study of the influence of the two-phonon terms
taking into account the p-p channel
on properties of the low-lying states is now in progress.

\section{Acknowledgments}

We are grateful to Prof. Ch.Stoyanov for valuable discussions and help.
A.P.S. and V.V.V. thank the hospitality of IPN-Orsay where a part
of this work was done.
This work is partly supported
by INTAS Fellowship grant for Young Scientists
(Fellowship Reference N YSF 2002-70), and
by IN2P3-JINR agreement.

\begin{appendix}
\section {}
The coefficients of the hamiltonian  (\ref{coup1}) are given by the
following expressions:

\begin{equation}
W^{\lambda ii^{\prime }}\left( \tau \right)=\sum_{k=1}^{N}
\left( \frac {D_M^{\lambda i k}\left( \tau \right)}
{\sqrt{2{\cal Y}_\tau ^{\lambda k i'}}} +
\frac {D_M^{\lambda i' k}\left( \tau \right)}
{\sqrt{2{\cal Y}_\tau ^{\lambda k i}}} \right),
\end{equation}
\begin{equation}
\Gamma_{jj'}^{\lambda i}\left( \tau \right)=\sum_{k=1}^{N}
\frac{f_{jj^{^{\prime }}}^{(\lambda k)}
v_{jj^{^{\prime }}}^{\left( -\right) }}
{\sqrt{2{\cal Y}_\tau ^{\lambda k i}}},
\end{equation}

where

\[
D_M^{\lambda i k}\left( \tau \right) =\left. \sum_{jj^{^{\prime }}}\right.
^\tau f_{jj^{^{\prime }}}^{(\lambda k)}u_{jj^{^{\prime
}}}^{\left( +\right) }\left( X_{jj^{^{\prime }}}^{\lambda i}+ Y_{jj^{^{\prime }}}^{\lambda i}
\right),
\]

\[
{\cal Y}_\tau ^{\lambda ki}=\frac{2\left( 2\lambda +1\right) ^2}{\left(
D_M^{\lambda ik}\left( \tau \right) \left( \kappa _0^{\left( M,k\right)
}+\kappa _1^{\left( M,k\right) }\right) +D_M^{\lambda ik}\left( -\tau
\right) \left( \kappa _0^{\left( M,k\right) }-\kappa _1^{\left( M,k\right)
}\right) \right) ^2},
\]

$v_{jj^{^{\prime }}}^{(-)}\,=\,u_ju_{j^{\prime }}\,-
\,v_jv_{j^{\prime }}$ $u_{jj^{^{\prime }}}^{(+)}=u_jv_{j^{^{\prime }}}+
\,v_ju_{j^{^{\prime }}}$.

In the above expressions, $f_{jj^{^{\prime }}}^{(\lambda k)}$ denotes the
single-particle radial
matrix elements \cite{ssvg02}:
\[
f_{j_1j_2}^{(\lambda k)}=u_{j_1}(r_k)u_{j_2}(r_k)i^\lambda \langle
j_1||Y_\lambda ||j_2\rangle,
\]
where $u_{j_1}(r_k)$ is the radial part of the HF single-particle wavefunction at the abscissas of
the $N$-point integration Gauss formula $r_k$.
$\kappa _0^{(M,k)}$ and $\kappa _1^{(M,k)}$ are defined by the Landau parameters as

\[
\left(
\begin{array}{c}
\kappa _0^{(M,k)} \\
\kappa _1^{(M,k)}
\end{array}
\right) =-N_0^{-1}\frac{Rw_k}{2r_k^2}\left(
\begin{array}{c}
F_0(r_k) \\
F_0^{\prime }(r_k)
\end{array}
\right).
%\label{}
\]

\section {}
The matrix elements $U_{\lambda _2i_2}^{\lambda _1i_1}(J i)$ have the following form:
\begin{eqnarray}
U_{\lambda _2i_2}^{\lambda _1i_1}(\lambda i) &=&(-1)^{\lambda
_1+\lambda _2+\lambda }\sqrt{(2\lambda _1+1)(2\lambda _2+1)}
\left. \sum_\tau \sum_{j_1j_2j_3}\right. ^\tau\times \\
&&\ \ \left({\Gamma_{j_1j_2}^{\lambda i}\left( \tau \right)}
\left\{
\begin{array}{ccc}
\lambda _1 & \lambda _2 & \lambda \\
j_2 & j_1 & j_3
\end{array}
\right\} \left( X_{j_2j_3}^{\lambda _2i_2} Y_{j_3j_1}^{\lambda _1i_1}
+ X _{j_3j_1}^{\lambda _1i_1} Y_{j_2j_3}^{\lambda _2i_2}\right) +\right. \nonumber \\
&&\ \ \Gamma_{j_1j_2}^{\lambda _1i_1}\left( \tau \right)
\left\{
\begin{array}{ccc}
\lambda _1 & \lambda _2 & \lambda \nonumber \\
j_3 & j_2 & j_1
\end{array}
\right\}\left( Y_{j_3j_1}^{\lambda _2i_2}Y _{j_2j_3}^{\lambda i}+
X_{j_2j_3}^{\lambda i}
X_{j_3j_1}^{\lambda _2i_2}\right) + \nonumber \\
&&\ \ \left. \Gamma_{j_1j_2}^{\lambda _2i_2}\left( \tau \right)
\left\{
\begin{array}{ccc}
\lambda _1 & \lambda _2 & \lambda \nonumber \\
j_1 & j_3 & j_2
\end{array}
\right\} \left(Y_{j_2j_3}^{\lambda _1i_1} Y _{j_3j_1}^{\lambda i}+
X _{j_3j_1}^{\lambda i}X_{j_2j_3}^{\lambda _1i_1}\right) \right).
%\label{}
\end{eqnarray}

\end{appendix}

%\newpage

\begin{table}[]
\caption[]{Energies and B(E2)-values for up-transitions to the first
$2^{+}$ states}
\begin{center}
\begin{tabular}{ccccccc} %\hline
Nucleus    & \multicolumn{3}{c} {Energy}       & \multicolumn{3}{c} {B(E2$\uparrow$)}    \\
           & \multicolumn{3}{c} {(MeV)}        & \multicolumn{3}{c} {(e$^2$b$^2$)}      \\
           & Exp. &  \multicolumn{2}{c}{Theory}& Exp. &  \multicolumn{2}{c}{Theory}      \\
           &      &  QRPA & 2PH                &      &   QRPA & 2PH                     \\
\hline
$^{124}$Sn & 1.13 &  1.92 &  1.03              & 0.1660$\pm$0.0040 &  0.177 & 0.151            \\
$^{126}$Sn & 1.14 &  1.96 &  1.30              & 0.10$\pm$0.03     &  0.149 & 0.133            \\
$^{128}$Sn & 1.17 &  2.08 &  1.48              & 0.073$\pm$0.006   &  0.111 & 0.100            \\
$^{130}$Sn & 1.22 &  2.37 &  1.73              & 0.023$\pm$0.005   &  0.064 & 0.058            \\
$^{132}$Sn & 4.04 &  4.47 &  4.03              & 0.14$\pm$0.06     &  0.136 & 0.129            \\
$^{134}$Sn & 0.73 &  1.65 &  1.34              & 0.029$\pm$0.006   &  0.016 & 0.015            \\
%\hline
\end{tabular}
\end{center}
\end{table}

\begin{table} []
\caption[]{Energies and B(E3)-values for up-transitions to the first
 $3^{-}$ states}
\begin{center}
\begin{tabular}{ccccccc}%\hline
Nucleus    & \multicolumn{3}{c} {Energy}       & \multicolumn{3}{c} {B(E3$\uparrow$)}    \\
           & \multicolumn{3}{c} {(MeV)}        & \multicolumn{3}{c} {(e$^2$b$^3$)}      \\
           & Exp. &  \multicolumn{2}{c}{Theory}& Exp. &  \multicolumn{2}{c}{Theory}      \\
           &      &  QRPA & 2PH                &      &   QRPA & 2PH                     \\
\hline
$^{124}$Sn & 2.60 &  3.64 & 3.25               & 0.073$\pm$0.010& 0.208  &  0.196         \\
$^{126}$Sn & 2.72 &  4.16 & 3.76               &                & 0.191  &  0.176         \\
$^{128}$Sn &      &  4.66 & 4.22               &                & 0.181  &  0.161         \\
$^{130}$Sn &      &  5.17 & 4.75               &                & 0.183  &  0.159         \\
$^{132}$Sn & 4.35 &  5.66 & 5.36               &                & 0.202  &  0.191         \\
$^{134}$Sn &      &  5.01 & 4.51               &                & 0.128  &  0.111         \\
%\hline
\end{tabular}
\end{center}
\end{table}

\begin{table}[]
\caption[]{ $(M_n/M_p)/(N/Z)$ ratios for the first $2^{+}$, $3^{-}$ states}
\begin{center}
\begin{tabular}{ccccccc} %\hline
 State    & $^{124}$Sn & $^{126}$Sn & $^{128}$Sn & $^{130}$Sn & $^{132}$Sn & $^{134}$Sn \\
\hline
 $2^+_1$   & 0.99       & 0.99      & 0.98       & 0.97       & 0.81       & 1.44        \\
 $3^-_1$   & 0.94       & 0.92      & 0.89       & 0.86       & 0.83       & 0.91        \\
%\hline
\end{tabular}
\end{center}
\end{table}

\end{document}